\documentstyle[12pt, aaspp4, tighten]{article}
\def \sun {$_{\scriptscriptstyle \odot}$}
\lefthead{FRYER}
\righthead{Black Hole Formation}
\begin{document}
\begin{center} {\em The Astrophysical Journal} in press
\end{center}
\vspace{1.cm}
\title{Mass Limits For Black Hole Formation}
\author{Chris L. Fryer}
\affil{Lick Observatory, University of California Observatories,
\\  Santa Cruz, CA 95064 \\ cfryer@ucolick.org}
\authoremail{cfryer@ucolick.org}

\begin{abstract}

We present a series of two-dimensional core-collapse supernova simulations 
for a range of progenitor masses and different input physics.  
These models predict a range of supernova energies 
and compact remnant masses.  In particular, we study two mechanisms 
for black hole formation:  prompt collapse and delayed collapse 
due to fallback.  For massive progenitors ($>$20\,M\sun), after a 
hydrodynamic time for the helium core (a few minutes to a few hours), 
fallback drives the compact object beyond the maximum neutron star mass 
causing it to collapse into a black hole.  With the current accuracy 
of the models, progenitors more massive than 40\,M\sun\, form black 
holes directly with no supernova explosion (if rotating, these black 
holes may be the progenitors of gamma-ray bursts).  We calculate the mass 
distribution of black holes formed, and compare these predictions 
to the observations, which represent a small biased subset of the 
black hole population.  Uncertainties in these estimates are 
discussed.

\end{abstract}


\section{Introduction}

As the number of massive compact accretors in X-ray binaries increases 
(McClintock \& Remillard 1986; Casares, Charles, \& Naylor 1992; Remillard, 
McClintock, \& Bailyn 1992; Bailyn et al. 1995; Filippenko, 
Matheson, \& Barth 1995; Remillard et al. 1996), 
so does the importance of understanding the formation of these 
stellar-mass black holes.  Although it has long been known that 
stellar-mass black holes could form from the collapse of massive 
stars (Oppenheimer \& Snyder 1939), theorists have yet to explain 
any details of black hole formation:  e.g. the number or mass 
distribution of the black holes formed.

This lack of progress in understanding black hole formation 
is a result of the difficulty in modeling the core collapse 
of massive stars.  Pursuit of the relevant physics of 
core-collapse supernovae has occupied theorists for three decades 
(see Bethe 1990).  The evidence suggesting that black holes form 
from stars with masses above 25\,M\sun continues to grow and includes:  
nucleosynthetic constraints (Maeder 1992; Kobulnicky \& Skillman 1997) 
and the formation of black hole X-ray binaries (Portegies Zwart, 
Verbunt, \& Ergma 1992; Ergma \& van den Heuvel 1998).  Not until the 
last decade, with the acceptance (and the successful 2D simulations) 
of the delayed neutrino-driven supernova mechanism 
(Wilson \& Mayle 1988; Herant et al. 1994; Burrows, Hayes, \& Fryxell 
1995; Janka \& M\"uller 1996; Fryer 1998), has it become possible 
for simulations of core collapse to make predictions on black hole 
formation.  Unlike the constraints from nucleosynthesis and from 
X-ray binary formation, core-collapse simulations provide direct 
evidence for black hole formation.  Black holes can form in core 
collapse either by direct collapse of a massive star or through 
fallback after a supernova explosion.  
In this paper, we outline the conditions required to 
produce black holes and apply these conditions to the results of 
core-collapse simulations.  From these simulations we can determine 
the number and mass distribution of black holes.

\section{Black Hole Formation}
To understand black hole formation, one must first understand 
the mechanism behind core-collapse supernovae.  The current paradigm 
is based upon an explosion driven by neutrino-energy deposition.  
A shock is produced as the inner core of a massive star collapses and 
bounces.  The shock stalls due to dissociation and neutrino losses but 
leaves behind an unstable entropy gradient.  This entropy gradient
initiates a convective layer at the edge of the stalled shock
which grows down to the proto-neutron star surface.
Neutrino heating drives the convection further as cool material
flows down to the proto-neutron star, heats via neutrino absorption
and rises and expands before it can lose its energy through
neutrino emission.  The outer edge of the convection layer is
bounded by an accretion shock as the star continues to collapse
on itself.  The ram pressure of the shock is given by:
\begin{equation}\label{eq:pshock}
P_{\rm shock}=
\frac{1}{2} \rho_{\rm S} v_{\rm ff}^2 =
\frac{\sqrt{2 G M_{\rm encl}} \dot{M_{\rm S}}}{8 \pi R_{\rm S}^{2.5}}
\end{equation}
where $G$ is the gravitational constant,
$v_{\rm ff}=\sqrt{2 G M_{\rm encl}/R_{\rm S}}$, $\rho_{\rm S}$,
$\dot{M_{\rm S}}$, and $M_{\rm encl}$ are, respectively, the free-fall
velocity, density, mass infall rate and enclosed mass just above the
shock radius ($R_{\rm S}$).  The pressure in the convective layer must
overcome this ram pressure to drive a successful explosion.

Once the convective layer begins to push the shock radius outward, 
the pressure from the shock ($P_{\rm shock}$) decreases, and an 
explosion is virtually inevitable (Bethe 1997).  However, if the 
shock pressure overcomes the pressure in the convective layer, 
its radius decreases, and it becomes even more difficult for the 
convective layer to overcome the ram pressure.  In these cases, 
the star collapses directly into a black hole\footnote{This does 
not preclude such a collapse from being observed.  If the star 
is rotating rapidly enough, it will form an accretion disk which 
can power a gamma-ray burst (Woosley 1993; 
MacFadyen \& Woosley 1999).}.  Unfortunately for supernova 
theorists, the most recent simulations find that massive cores 
straddle the fine line between explosion and collapse 
(Wilson \& Mayle 1988; Miller, Wilson, \& Mayle 1993; 
Herant et al. 1994; Burrows, Hayes, \& Fryxell 1995; Janka 
\& M\"uller 1996; Mezzacappa et al. 1998; Messer et al. 1998; 
Fryer 1998).  Because core-collapses straddle this line, their 
ultimate outcome depends sensitively upon the implementation of 
the physics (e.g. equation of state, neutrino transport, general 
relativity) as well as upon the progenitor (e.g. progenitor 
mass or rotation).  Burrows \& Goshy (1993) stressed 
the importance of the mass infall rate for the success or 
failure of a supernova explosion.  This is directly related 
to the progenitor mass, because, at any given time after collapse, 
the infall rate increases with increasing progenitor mass (Fig. 1).
As the mass infall rate increases, the shock pressure increases 
(Eq. \ref{eq:pshock}), and the convective layer must have more 
energy to explode.  The large difference between 15 and 25\,M\sun 
progenitors is due to differences in the iron core mass of these 
models (Weaver \& Woosley 1993, 1996; Timmes, Woosley, \& Weaver 
1996).   Above some progenitor star mass, all stars will directly 
collapse to black holes, forming black holes of mass equal to their 
progenitor.

But even those stars which explode may form black holes.  As the 
supernova shock travels outward, it decelerates (Sedov 1959): 
\begin{equation}
v_{shock} \propto t^{\frac{\omega-3}{5-\omega}},
\end{equation}
where $\omega$ is given by the density structure of the medium through
which the shock travels ($\rho \propto r^{-\omega}$).  Some of the 
expanding material may decelerate below the escape velocity and 
fall back onto the neutron star (Herant \& Woosley 1994, 
Woosley \& Weaver 1995).  If this material pushes the neutron star 
above the maximum neutron star mass limit, a black hole is formed.  
In this manner, the core-collapse of a massive star can produce both 
a supernova and a black hole.  The mass of these black holes depends 
upon the amount of fallback which ultimately produces a range of black 
hole masses.

Thus, for core collapse models, we can define three regimes of 
compact object formation:  a) low mass, core-collapse stars drive 
strong explosions with little fallback and produce neutron stars, 
b) moderate mass stars produce explosions, but the fallback is 
sufficient to form black holes, and c) high mass stars are unable 
to launch shocks and collapse directly to black holes.  The 
question for core-collapse theorists, then, is to determine 
the limits for these regimes.

\section{Core-Collapse Simulations}

For our simulations, we use a code originally described in 
Herant et al. (1994).  This code models the core collapse continuously 
from collapse through bounce and ultimately to explosion.  The 
neutrino transport is mediated by a crude, single energy flux-limiter.  
Beyond a critical radius, $\tau < 0.3$, a simple ``light-bulb'' 
approximation for the neutrinos is invoked which assumes that any material 
beyond that radius is bathed by an isotropic flux equal to the neutrino 
flux escaping that radius.  For our simulations, we have raised this 
radius to $\tau < 0.1$ (which modified the kinetic energies 
by 10\%), and we also removed the neutrino/electron scattering 
opacity\footnote{This can have large effects.  See Swesty (1998)}.  
The angular resolution has been improved to roughly $1^{\circ}$.
To this code, we have added spherically symmetric general 
relativity and a more sophisticated flux limiter (Fryer et al. 1999).  
The advantage of this code is 
that it models the supernova explosion from collapse through 
bounce without the need to set up a new grid.  In addition, 
all but the inner $0.001-0.004\,$M\sun is modeled in 2-dimensions, 
avoiding any problems that might arise from constructing an inner 
boundary.  The drawback of this code is its single-energy flux-limited 
neutrino transport.  Because massive cores straddle the line between a 
supernova explosion and a direct collapse into black hole, the details 
of all the input physics (e.g. equation of state, general relativity) 
are important, including the algorithm for neutrino transport 
(Janka \& M\"uller 1996, Mezzacappa et al. 1998, Messer et al. 1998).  
We will come back to the uncertainties in the physics in the 
next section.

First, however, let's review the results of our simulations.  
Table 1 summarizes the entire set of simulations, using 3 
progenitor masses (15\,M\sun,25\,M\sun,40\,M\sun) both with 
and without the effects of general relativity.  The 
``standard'' models\footnote{These models are the most physical 
of our models.  We do not artificially alter the neutrino flux 
and include the effects of general relativity.} are given in bold-face. 
In addition, because Mezzacappa et al. (1998) found that their more 
detailed neutrino transport lead to lower neutrino energies and 
luminosities (by roughly 10\%), we have
run a set of models where the neutrino energies are artificially 
lowered by 20\%.  This lowers the luminosity by 20\%.  Since 
the neutrino opacity is proportional to the square of the neutrino 
energy, it lowers the amount of neutrino heating by an additional 
40\%.  This lowered neutrino run leads to energies and luminosities 
which are lower than those of Mezzacappa et al. (1998) and, combined 
with our ``standard'' runs, brackets their results.  If the differences 
in the models are simply caused by differences the neutrino energy, 
by lowering the neutrino energies by 20\%, our 15 M\sun model should have 
fizzled along with the 15 M\sun models of Mezzacappa et al. (1998).
In figure 2, note that our mean neutrino energies and luminosities 
are indeed lower than those of Mezzacappa et al. (1998), yet 
from Table 1, we see that we still get an explosion.  Clearly, 
the differences in the mean neutrino energies can not explain 
all the differences in the simulation.  However, our low neutrino 
run allows us to estimate the sensitivity of the core-collapse 
simulations to the neutrino transport.

The trends in the compact remnant masses and explosion energies 
can be understood by comparing the shock pressure to the pressure 
in the convective region.  By lowering the neutrino energies, 
there is less heating and the convective layer has less pressure.  
It therefore takes longer for the convective layer to overcome the 
ram pressure.  The collapsed core accumulates more mass, and generates 
less energetic explosions.  Although the increased effective mass using 
general relativisitic gravity leads to a faster (by 10 ms) 
bounce, the lower heating rate (due to both the time dilation and 
the redshift of the neutrinos) leads to weaker convection, and 
a later explosion.  

The differences in the ram pressure for different progenitors 
also explains the varying results for the massive progenitors.  
For the 15\,M\sun, the mass infall rate (Fig. 1), and hence 
ram pressure, is 5 times lower just 100\,ms past bounce.
It is not surprising, then, that the 15\,M\sun model explodes much 
faster than its more massive counterparts.  Figure 3 shows the 
evolution of the 15\,M\sun and 25\,M\sun models with time 
(in the standard models).  
The convective layer in the 15\,M\sun model quickly overcomes 
the ram pressure and launches an explosion 140\,ms past bounce.  
The 25\,M\sun model takes nearly 100\,ms longer to explode.  Since 
the infall rate of the 25\,M\sun and 40\,M\sun progenitors 
do not differ significantly until 300\,ms past bounce, it is 
not surprising that these simulations give similar answers.  However, 
the 40\,M\sun model teeters on the edge of direct collapse (lowering 
the neutrino energy produces no explosion, and at the end of 
the simulation, the accretion shock radius is decreasing).  
For our simulations, the 40\,M\sun progenitor roughly marks the 
dividing line between supernova explosion and direct collapse.

But what about the lower progenitor mass-limit for black hole formation due 
to fallback?  The explosion energy in Table 1 does not include 
the binding energy of the ejected material, and this binding energy 
must be overcome to drive an explosion.
We can roughly estimate the fallback by assuming that only 
the outer layers with total binding energy less than the 
explosion energy are actually ejected (Table 1).  From 
the models of Woosley \& Weaver (1995), we can calculate 
the energy required to eject all but the inner 3\,M\sun 
core (Fig. 4).  If the explosion energy is less than this amount, 
the compact remnant will accrete beyond 3\,M\sun and will 
collapse to a black hole.  Note that our explosion energies 
decrease with increasing progenitor mass, whereas the binding 
energy of the star increases with increasing mass.  These 
two effects limit the ``neutron star-fallback black hole'' 
transition mass to a narrow range (18-25\,M\sun).

However, one must be careful about the definitions of these 
energies.  The explosion energy given here is computed by calculating 
the difference (before and after the explosion) between the sum of 
kinetic + internal - potential energies of the material beyond the 
core mass.  As a lower limit, this energy must overcome the binding 
energy of the star to avoid the formation of a black hole.  
In fact, it must be greater to actually produce an energetic 
supernova.  In practice, some of the energy from our simulations 
goes into the kinetic energy of the explosion (KE$_\infty$ of 
Woosley \& Weaver 1995), and the amount of fallback predicted 
in Table 1 is a lower limit.  Note that, especially in observational 
papers, the term ``explosion'' energy is used to mean KE$_\infty$.  
To extract KE$_\infty$ from the explosion energies given in Table 1, 
one must subtract the binding energy of the material which is ejected.

For example, the mass of the progenitor of supernova 1987A is 
thought to be $\sim$\,20M\sun, and yet its energy at infinity was roughly 
$10^{51}$\,ergs (Woosley 1988) which corresponds to over  
$2 \times 10^{51}$\,ergs of core-collapse explosion energy (Table 1).
Recently, a number of supernovae with low Nickel ejecta have 
been discovered:  SN 1997D (Turatto et al. 1997) and 
SN 1994D (Sollerman, Cumming \& Lundqvist 1998).  
Turatto et al. (1997) were able to match the spectra and light curves 
of 1997D with a 26\,M\sun progenitor and a low KE$_\infty = 
0.4\times 10^{51}$\,ergs explosion.  However, to eject all but the inner 
$\sim 2.1$\,M\sun\, with an energy at infinity of 
$0.4\times 10^{51}$\,ergs requires roughly $1.4\times 10^{51}$\,ergs 
explosion energy, twice the value for our 25\,M\sun\, model (Table 1).  
Clearly, the details of the core-collapse model must be understood better.  

\section{Implications and Uncertainties}

Core-collapse simulations can now place rough limits on 
black hole formation:  stars more massive than $\sim 25$\,M\sun 
will eventually collapse to form a black hole, and those more 
massive than 40\,M\sun will not produce a supernova explosion.  
Assuming a Scalo (1986) initial mass function 
($\alpha_{\rm IMF} = 2.7$), the ratio of black holes to 
neutron stars in the Galaxy is 16\% (9.3\% from fallback, 
7.5\% from direct collapse).  This number does not include 
those black holes formed from hypercritical accretion onto 
neutron stars in binaries (Bethe \& Brown 1998; Fryer \& Woosley 
1998; Fryer, Woosley, \& Hartmann 1999) which may double this number.

From Table 1, we see that the black hole mass should range 
from 3-15\,M\sun for progenitors less massive than 40\,M\sun.  
Beyond 40\,M\sun, the final black hole mass could be as large 
as its progenitor.  But the progenitor mass depends 
sensitively on the implementation of winds and binary effects.  
Bailyn et al. (1998) have suggested that the masses of black 
holes cluster around $\sim 7$\,M\sun.  The black holes 
that have been measured are all in X-ray binaries.  It is 
likely that the progenitors of these black holes lost most 
of their hydrogen envelope in a common envelope evolution.  
The loss of this hydrogen envelope will not significantly change the 
core, nor the results of the core-collapse simulations 
dramatically, but it will change the amount of material 
which can fall back onto the core.  26\,M\sun and 45\,M\sun 
stars have helium cores of mass 10\,M\sun, 20\,M\sun respectively.  
Given that some of the helium core mass will be lost to winds, 
and further mass will be ejected in the supernova explosion, 
these black holes should, on the average, be less massive 
than their single-star counterparts, but in any case, 
these black holes should have a range of masses:  3-15\,M\sun.  
With the current data, this range fits the data as well as the 
Bailyn et al. (1998) single mass value.  If the data improves and 
exhibits no range whatsoever, an important piece of the black hole 
puzzle is still missing.  A range of black hole masses 
will support our outline of black hole formation.

However, the sensitivity of the core-collapse simulations to the 
implementation of the physics, both in the core collapse and the 
progenitor models, strongly argues for caution in any of these 
claims.  Simply by lowering the mean neutrino energy by 20\% decreases 
the resultant explosion energy by over a factor of 2.  This lowers the 
fallback black hole mass limit to roughly 15\,M\sun, increasing 
the fraction of black holes from 16\% to 52\%!  Both the approximations 
used to calculate the neutrino transport in multi-dimensions (e.g. 
a single-energy flux-limited diffusing scheme in our case) and 
the errors in neutrino cross-sections and emissivities can lead to 
these 20\% errors.  The uncertainties in the core-collapse models 
are not limited to neutrinos.  The distance the initial bounce 
travels before stalling (and hence the strength of the shock ram 
pressure) depends upon the equation of state.  Unfortunately, the 
behavior of gas at nuclear densities depends on uncertain particle 
physics.  Comparisons of 1-dimensional simulations with no convection 
(which do not explode) with 2-dimensional simulations (which do 
explode) show that convection can help an explosion, but the 
quantitative effect of convection can not be determined until 
full 3-dimensional simulations are simulated.  

In addition, the structure of the progenitor effects the outcome 
of the collapse, and much of the physics in the progenitor model 
remains uncertain:  opacities, implementation of convection, and 
mass loss from winds.  Rotation will also effect the supernova models.  
M\"onchmeyer (1991) has argued that rotation weakens the bounce 
and, hence, the explosion, but Yamada \& Sato (1994) have found 
that the asymmetric neutrino emission found by Janka \& 
M\"onchmeyer (1989) can drive stronger convection, allowing 
more neutrino energy to convert to kinetic energy, ultimately 
driving a stronger explosion. However, the good agreement of the 
general picture of black hole formation does imply (but does not prove) 
that the solution has indeed moved to a study of the details.  
As the uncertainties of the physics are better understood, the 
reliability of the core-collapse predictions of black hole formation 
will increase. 

\acknowledgements
This research has been supported by NASA (NAG5-2843, MIT SC A292701, 
and NAG5-8128), the NSF (AST-97-31569), and the US DOE ASCI Program 
(W-7405-ENG-48).  It is a pleasure to thank Stan Woosley for 
encouragement, advice, and access to his models.  Conversations with 
Thomas Janka, Adam Burrows, and especially Tony Mezzacappa contributed 
to this work.  I'd also like to thank Alex Heger and Aimee Hungerford 
for their helpful comments.

\begin{deluxetable}{lcccccc}
\tablewidth{35pc}
\tablecaption{Core Masses, Explosion Energies, and Ejecta Masses}
\tablehead{ \colhead{Model} & \colhead{M$_{\rm Core}$\tablenotemark{a}}  
& \colhead{M$_{\rm Remnant}$\tablenotemark{b}}  
& \colhead{Energy\tablenotemark{c}}  
& \multicolumn{3}{c}{Mass ejected (M\sun)} \\
 &  (M\sun) & (M\sun) & ($10^{51}$\,Erg) & $Y_e < 0.4$ & $Y_e < 0.45$ 
& $Y_e < 0.49$}

\startdata
15\,M\sun Newtonian & 1.1 & 1.1 & 3.0 & 0.15 & 0.19 & 0.24 \nl
{\bf 15\,M\sun GR}\tablenotemark{d} & {\bf 1.2} & {\bf 1.4} & {\bf 2.5} & 
{\bf 0.07} & {\bf 0.13} & {\bf 0.17} \nl
15\,M\sun GR-low $\nu$ & 1.4 & 2.2 & 0.1 & -\tablenotemark{e} & - & - \nl
25\,M\sun Newtonian & 1.3 & 1.3 & 2.2 & 0.25 & 0.30 & 0.54 \nl
{\bf 25\,M\sun GR} & {\bf 1.4} & {\bf 5.2} & {\bf 0.6} 
& -\tablenotemark{e} & - & - \nl
25\,M\sun GR-low $\nu$ & 1.6 & 25 & 0. & - & -  & -  \nl
{\bf 40\,M\sun GR} & {\bf 1.6} & {\bf 12.9} & {\bf 0.6} 
& -\tablenotemark{e} & - & - \nl
40\,M\sun GR-low $\nu$ & $>1.6$ & 40 & 0. & - & - & -  \nl

\tablenotetext{a}{The core mass assumes no fallback.  All masses 
are the baryonic mass.}
\tablenotetext{b}{The remnant mass after fallback estimated 
by assuming only material with binding energy less than the 
supernova energy is actually ejected and that no mass is 
lost from winds.} 
\tablenotetext{c}{The explosion energy is computed by calculating 
the difference (before and after the explosion) between the sum of 
kinetic + internal - potential energies of the material beyond the 
core mass.  This energy must overcome the binding energy of the 
star to avoid the formation of a black hole.}
\tablenotetext{d}{The results in bold-faced are the ``most-likely'' 
given the current sophistication of the models.  The different 
variations in the results, however, give some idea of the 
range in these results.}
\tablenotetext{e}{The low energy 15\,M\sun, 25\,M\sun\, runs 
as well as the 40\,M\sun ``standard'' run will not eject 
significant amounts of neutron rich material unless it is 
carried out by convection.}
\enddata
\end{deluxetable}
\clearpage

\begin{figure}
\plotfiddle{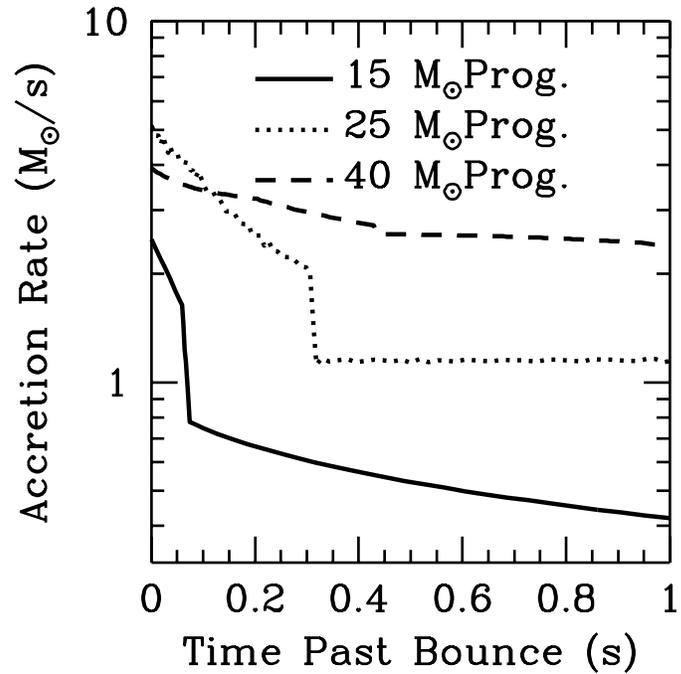}{7in}{0}{100}{100}{-165}{0}
\caption{Mass infall rates for a 3 separate progenitor masses: 
15,25,40\,M\sun.  The mass infall rate for the 15\,M\sun progenitor 
drops to 1/5th that of the 25 and 40\,M\sun models in 100\,ms.  
This allows it to explode sooner, leaving behind a smaller core.  
The infall rates of the 25 and 40\,M\sun progenitors stay 
roughly the same for 300\,ms past bounce, and hence their 
explosion energies are similar.}
\end{figure}

\begin{figure}
\plotfiddle{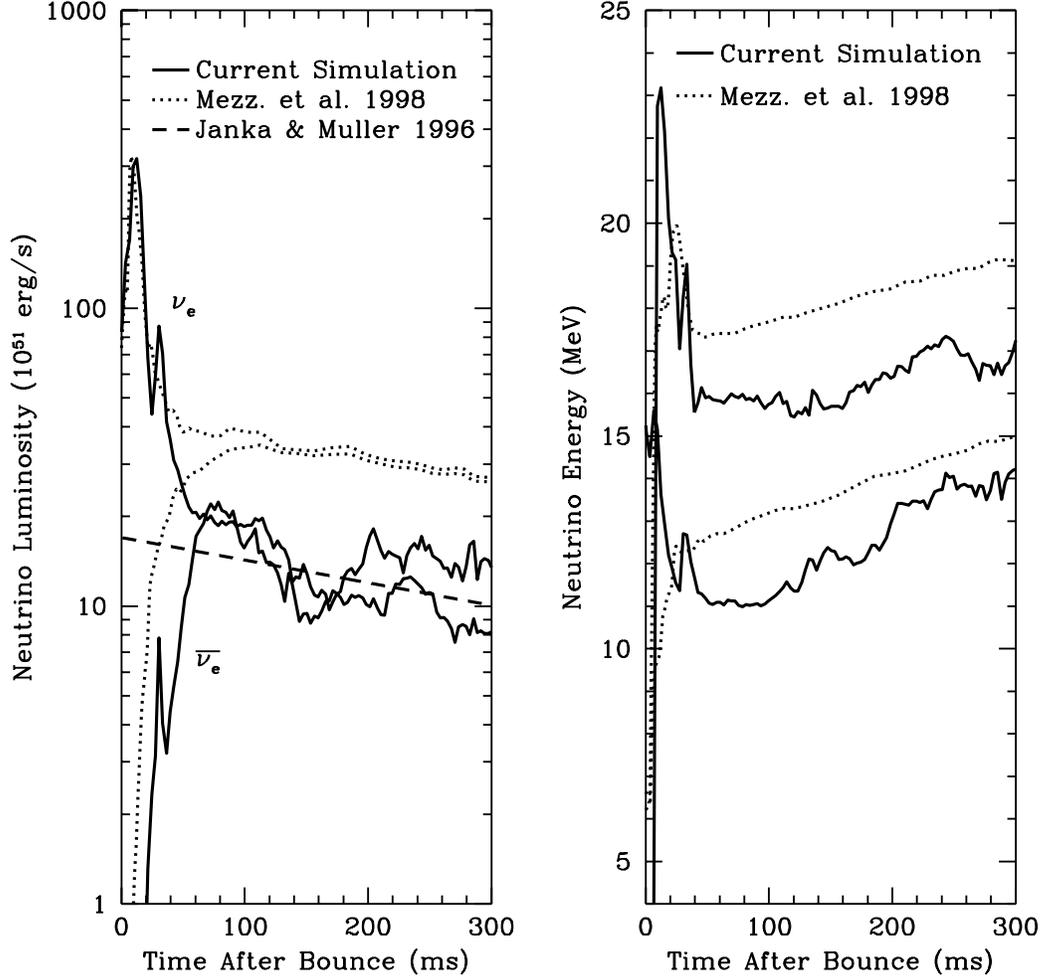}{7in}{0}{70}{70}{-220}{0}
\caption{(a) Electron neutrino and anti-electron neutrino 
luminosities and (b) energies for our 15\,M\sun
run with lowered neutrino energy and the  15\,M\sun simulation 
of Mezzacappa et al. (1998).  Note that our luminosity and energy 
is less than or equal to theirs, yet our simulation explodes and 
theirs collapses directly to a black hole.  We have 
also plotted the lowest neutrino luminosity from Janka \& M\"uller (1996) 
which leads to a supernova explosion in 2-dimensions.  
Their luminosity is also lower than that of Mezzacappa et al. (1998).}
\end{figure}

\begin{figure}
\plotfiddle{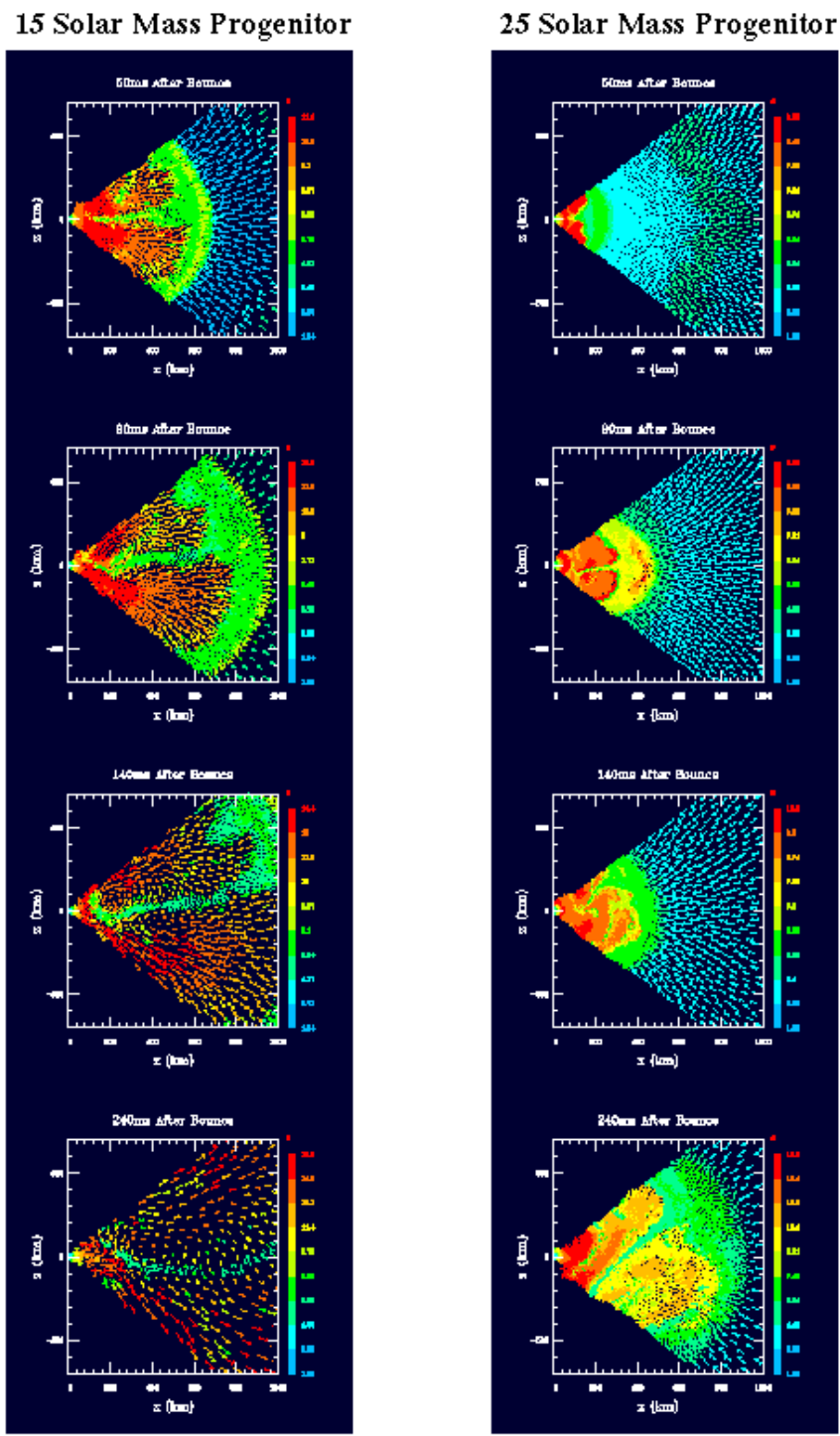}{7in}{0}{70}{70}{-210}{-10}
\caption{Snapshots of the evolution of both a 15 and 25\,M\sun 
core collapse (top to bottom: 50, 90, 140, 240\,ms).  
The 15\,M\sun model has launched a strong explosion after 140\,ms.  
It takes the 25\,M\sun progenitor nearly 100\,ms longer to develop 
such an explosion.  The color codes entropy with blue and red 
indicating limiting entropies of roughly 1,10$k_{\rm B}$ per 
nucleon respectively.  The vectors indicate the strength and 
direction of the velocities.}
\end{figure}

\begin{figure}
\plotfiddle{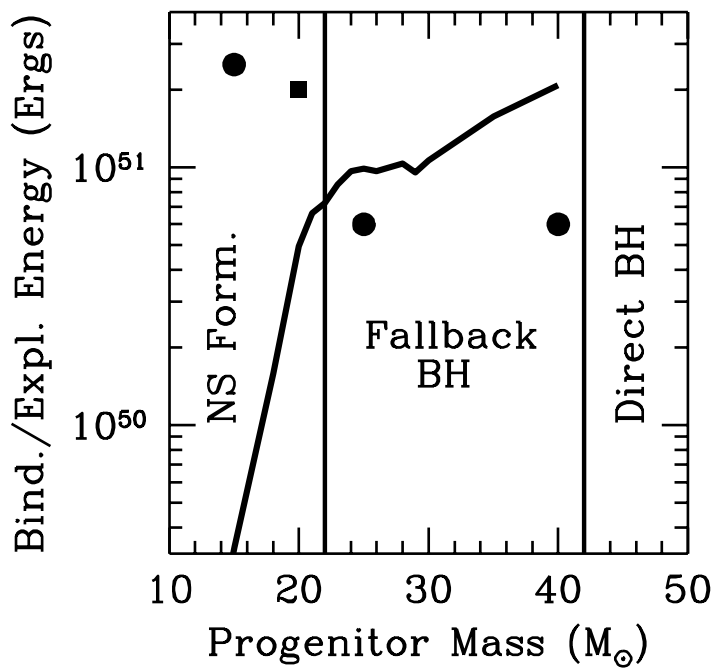}{7in}{0}{100}{100}{-165}{0}
\caption{Binding energy (solid line) and explosion energy 
(dots) vs. mass of progenitor.  This binding energy includes 
all but the inner 3\,M\sun core of the star.  If the explosion 
energy is less than the binding energy, the compact remnant 
will exceed 3\,M\sun and collapse to form a black hole.  
The explosion energy drops and the binding energy rises with 
increasing progenitor mass, their net effect is to create 
a fairly narrow range of uncertainty in the transition 
mass from neutron star formation and black hole formation 
from fallback.  For reference, supernovae 1987A is placed 
on the figure (square).}
\end{figure}

\end{document}